\documentclass[aps,prc,twocolumn,superscriptaddress,showpacs,nobalancelastpage]{revtex4}

\usepackage{graphicx}
\usepackage{dcolumn}
\usepackage{bm}

\begin{document}


\title{The $\vec{\gamma} p \rightarrow K^+ \Lambda$ and 
$\vec{\gamma} p \rightarrow K^+ \Sigma^0$ reactions at forward angles 
with photon energies from 1.5 to 2.4 GeV} 

\author{M. Sumihama}
  \affiliation{Research Center for Nuclear Physics, Osaka University, Ibaraki, Osaka 567-0047, Japan}
\author{J.K. Ahn}
  \affiliation{Department of Physics, Pusan National University, Busan 609-735, Korea}
\author{H. Akimune}
  \affiliation{Department of Physics, Konan University, Kobe, Hyogo 658-8501, Japan}
\author{Y. Asano}
  \affiliation{Synchrotron Radiation Research Center, Japan Atomic Energy Research Institute, Sayo, Hyogo 679-5198 , Japan}
\author{C. Bennhold}
\affiliation{The George Washington University, Washington, D.C., 20052, USA}
\author{W.C. Chang}
  \affiliation{Institute of Physics, Academia Sinica, Taipei, Taiwan 11529, Republic of China}
\author{T. Corthals}
  \affiliation{Department of Subatomic and Radiation Physics,  
  Ghent University, Proeftuinstraat 86, B-9000 Gent, Belgium}
\author{S. Dat\'e}
  \affiliation{Japan Synchrotron Radiation Research Institute, Sayo, Hyogo 679-5198, Japan}
\author{H. Ejiri}
  \affiliation{Japan Synchrotron Radiation Research Institute, Sayo, Hyogo 679-5198, Japan}
\author{H. Fujimura}
  \affiliation{Department of Physics, Kyoto University, Kyoto 606-8502, Japan} 
\author{M. Fujiwara}
  \affiliation{Research Center for Nuclear Physics, Osaka University, Ibaraki, Osaka 567-0047, Japan}
  \affiliation{Kansai Photon Science Institute, Japan Atomic Energy Agency, Kizu,
Kyoto 619-215, Japan}
\author{M. Guidal}
\affiliation{Institut de Physique, Nucl\'{e}aire d'Orsay, 91406 Orsay, France}
\author{K. Hicks}
  \affiliation{Department of Physics And Astronomy, Ohio University, Athens, Ohio 45701}
\author{T. Hotta}
  \affiliation{Research Center for Nuclear Physics, Osaka University, Ibaraki, Osaka 567-0047, Japan}
\author{K. Imai}
  \affiliation{Department of Physics, Kyoto University, Kyoto 606-8502, Japan} 
\author{T. Ishikawa}
  \affiliation{Laboratory of Nuclear Science, Tohoku University, Sendai, Miyagi 982-0826, Japan}
\author{T. Iwata}
  \affiliation{Department of Physics, Yamagata University, Yamagata 990-8560, Japan}
\author{H. Kawai}
  \affiliation{Department of Physics, Chiba University, Chiba 263-8522, Japan}
\author{Z.Y. Kim}
  \affiliation{School of Physics, Seoul National University, Seoul, 151-747, Korea}
\author{K. Kino}
  \affiliation{Research Center for Nuclear Physics, Osaka University, Ibaraki, Osaka 567-0047, Japan}
\author{H. Kohri}
  \affiliation{Research Center for Nuclear Physics, Osaka University, Ibaraki, Osaka 567-0047, Japan}
\author{N. Kumagai}
  \affiliation{Japan Synchrotron Radiation Research Institute, Sayo, Hyogo 679-5198, Japan}
\author{S. Makino}
  \affiliation{Wakayama Medical College, Wakayama, Wakayama 641-8509, Japan}
\author{T. Mart}
  \affiliation{Departemen Fisika, FMIPA, Universitas Indonesia, Depok 16424, Indonesia}
\author{T. Matsumura}
  \affiliation{Research Center for Nuclear Physics, Osaka University, Ibaraki, Osaka 567-0047, Japan}
\author{N. Matsuoka}
  \affiliation{Research Center for Nuclear Physics, Osaka University, Ibaraki, Osaka 567-0047, Japan}
\author{T. Mibe}
  \affiliation{Research Center for Nuclear Physics, Osaka University, Ibaraki, Osaka 567-0047, Japan}
\author{M. Miyabe}
  \affiliation{Department of Physics, Kyoto University, Kyoto 606-8502, Japan} 
\author{Y. Miyachi}
  \affiliation{Department of Physics, Tokyo Institute of Technology, Tokyo 152-8551, Japan}
\author{M. Morita}
  \affiliation{Research Center for Nuclear Physics, Osaka University, Ibaraki, Osaka 567-0047, Japan}
\author{N. Muramatsu}
  \affiliation{Research Center for Nuclear Physics, Osaka University, Ibaraki, Osaka 567-0047, Japan}
\author{T. Nakano}
  \affiliation{Research Center for Nuclear Physics, Osaka University, Ibaraki, Osaka 567-0047, Japan}
\author{M. Niiyama}
  \affiliation{Department of Physics, Kyoto University, Kyoto 606-8502, Japan} 
\author{M. Nomachi}
  \affiliation{Department of Physics, Osaka University, Toyonaka, Osaka 560-0043, Japan}
\author{Y. Ohashi}
  \affiliation{Japan Synchrotron Radiation Research Institute, Sayo, Hyogo 679-5198, Japan}
\author{T. Ooba}
  \affiliation{Department of Physics, Chiba University, Chiba 263-8522, Japan}
\author{H. Ohkuma}
  \affiliation{Japan Synchrotron Radiation Research Institute, Sayo, Hyogo 679-5198, Japan}
\author{D.S. Oshuev}
  \affiliation{Institute of Physics, Academia Sinica, Taipei, Taiwan 11529, Republic of China}
\author{C. Rangacharyulu}
  \affiliation{Department of Physics and Engineering Physics, University of Saskatchewan, Saskatoon, Saskatchewan, Canada, S7N 5E2} 
\author{A. Sakaguchi}
  \affiliation{Department of Physics, Osaka University, Toyonaka, Osaka 560-0043, Japan}
\author{T. Sasaki}
  \affiliation{Department of Physics, Kyoto University, Kyoto 606-8502, Japan} 
\author{P.M. Shagin}
  \affiliation{School of Physics and Astronomy, University of Minnesota, Minneapolis, Minnesota 55455}
\author{Y. Shiino}
  \affiliation{Department of Physics, Chiba University, Chiba 263-8522, Japan}
\author{H. Shimizu}
  \affiliation{Laboratory of Nuclear Science, Tohoku University, Sendai, Miyagi 982-0826, Japan}
\author{Y. Sugaya}
  \affiliation{Department of Physics, Osaka University, Toyonaka, Osaka 560-0043, Japan}
\author{H. Toyokawa}
  \affiliation{Japan Synchrotron Radiation Research Institute, Sayo, Hyogo 679-5198, Japan}
\author{A. Wakai}
  \affiliation{Akita Research Institute of Brain and Blood Vessels, Akita 010-0874, Japan}
\author{C.W. Wang}
  \affiliation{Institute of Physics, Academia Sinica, Taipei, Taiwan 11529, Republic of China}
\author{S.C. Wang}
  \affiliation{Institute of Physics, Academia Sinica, Taipei, Taiwan 11529, Republic of China}
\author{K. Yonehara}
  \affiliation{Illinois Institute of Technology, Chicago, Illinois 60616, USA}
\author{T. Yorita}
  \affiliation{Japan Synchrotron Radiation Research Institute, Sayo, Hyogo 679-5198, Japan}
\author{M. Yosoi}
  \affiliation{Research Center for Nuclear Physics, Osaka University, Ibaraki, Osaka 567-0047, Japan}
\author{R.G.T. Zegers}
  \affiliation{National Superconducting Cyclotron Laboratory (NSCL), Michigan State University, Michigan 48824-1321}
\collaboration{The LEPS collaboration}
  \noaffiliation

\date{\today}

\begin{abstract}
Differential cross sections and photon beam asymmetries for the 
 $\vec{\gamma} p \rightarrow K^+ \Lambda$ and 
$\vec{\gamma} p \rightarrow K^+ \Sigma^0$ reactions 
have been measured in the photon energy range from 1.5 GeV to 2.4 GeV 
and in the angular range from $\Theta_{cm} = 0^\circ$ to 60$^\circ$ 
of the $K^+$ scattering angle in the center of mass system at 
the SPring-8/LEPS facility. 
The photon beam asymmetries for both the reactions have been found to 
be positive and to increase with the photon energy. 
The measured differential cross sections agree 
with the data measured by the CLAS collaboration 
at $\cos\Theta_{cm}<$0.9 within the experimental uncertainties, 
but the discrepancy with the SAPHIR data for the $K^+\Lambda$ reaction 
is large at $\cos\Theta_{cm}>$0.9. 
In the $K^+\Lambda$ reaction, the resonance-like structure 
found in the CLAS and SAPHIR data at $W=1.96$ GeV is confirmed. 
The differential cross sections at forward angles suggest a strong 
$K$-exchange contribution in the t-channel for the $K^+ \Lambda$ 
reaction, but not for the $K^+ \Sigma^0$ reaction. 

\end{abstract}

\pacs{13.60.Le, 14.20.Gk, 25.20.Lj}

\maketitle

\section{INTRODUCTION}

By studying the excited states of baryons, deeper insight can be gained 
into their structure. The excited spectrum of baryons contains signatures 
stemming from the constituents at a more fundamental level. 
Experimental information on nucleon resonances ($N^*$ and $\Delta^*$) has 
been obtained primarily from the studies of the $\pi N$ and $N$($\gamma$, 
$\pi$) reactions. 
In spite of valuable information on resonances in pionic channels,  
a search of intermediate resonances might be limited in these reactions. 
This problem is addressed by recent calculations in constituent
quark models~\cite{Capstick1,Capstick2}. A considerably large number of 
nucleon resonances are predicted by theoretical calculations compared with 
those observed in the pionic reactions.  
The nucleon resonances predicted but not yet found 
are referred as `missing resonances'.     
Quark model studies suggest some of these missing resonances may 
couple to strange channels, such as $K\Lambda$ and $K\Sigma$ 
channels~\cite{Capstick2}. $\Lambda$ and $\Sigma^0$ hyperons have the isospins
 of 0 and 1, respectively. Accordingly, intermediate states of $K^+\Lambda$ 
have the isospin $\frac{1}{2}$ ($N^*$ only) whereas intermediate states of 
$K^+\Sigma^0$ can have both the isospins of $\frac{1}{2}$ and  $\frac{3}{2}$ 
($N^*$ and $\Delta^*$). It is very interesting to study the 
$\gamma p \rightarrow K^+ \Lambda$ and 
$\gamma p \rightarrow K^+ \Sigma^0$ reactions 
to further our understanding of the role that nucleon resonances 
play in non-pionic reactions.

Measurements of the total cross section for the $\gamma p
\rightarrow K^+ \Lambda$ 
reaction at ELSA/SAPHIR~\cite{saphirk+} showed a new resonance-like 
structure around W=1900 MeV ($E_\gamma$=1.5 GeV), where W is the total 
energy in the center of momentum system.  
The resonance-like structure was theoretically studied by 
connecting with missing resonances, like the $D_{13}$(1900), 
which were predicted to couple strongly to the $K\Lambda$ 
channel~\cite{MB, MAID, Janssen, Janssennew}.  
These theoretical calculations were performed 
in a tree-level effective-Lagrangian approach. 
The well-established resonances $S_{11}$(1650), 
$P_{11}$(1710) and $P_{13}$(1720) in an s-channel and  $K^*$ and $K_1$ 
in a t-channel were included~\cite{MB,MAID,Janssen,Janssennew}.   
There still remains some controversy in the theoretical description 
of the $K^+$ photoproduction, because of ambiguities from 
the choice of the resonances included, their meson-hadron couplings 
and form factors at hadronic vertices, and the treatment of the 
non-resonant background term. 
In particular, it has been found that the extracted resonance couplings
are greatly influenced depending on which background model~\cite{Janssen,
Saghai3,Janssennew2} is chosen. 
Thus, caution is advised in drawing conclusions on possible 
resonance structures. 

It is important to measure additional observables and obtain 
accurate experimental data over a wide kinematical range 
for developing of theoretical models and for improving  
our knowledge of $K^+$ photoproduction.  
Cross sections and recoil polarizations have been obtained at 
JLAB/CLAS~\cite{CLAS} and SAPHIR~\cite{saphirk+,saphirnew}. 
The cross section measurements at CLAS~\cite{CLAS} suggest that  
the resonance-like structure near 1900 MeV has more than one component 
by examining yields at different $K^+$ scattering angles. 
Recently, comprehensive measurements for 
the $\gamma p \rightarrow K^+ \Lambda$ 
and $\gamma p \rightarrow K^+ \Sigma^0$ reactions extending to 
$E_{\gamma}$=2.6 GeV at SAPHIR have been reported~\cite{saphirnew}. 
In addition, measurements of the photon beam asymmetry~\cite{KPRL} and 
the transferred polarization in kaon electroproduction~\cite{Carman} 
help to further define the $K^+$-photoproduction mechanism. 

Furthermore, to calculate the cross sections of 
hypernuclear electroproduction, it is helpful to know the precise 
cross sections of the elementary reaction of kaon photoproduction. 
Therefore an improvement of the experimental data is also quite important
 from the view point of hypernuclear physics. 

The contribution of $t$-channel meson exchange is expected to become 
large at forward angles above the resonance region at $E_\gamma>$2 GeV.    
Mesons exchanged in kaon photoproduction are $K$, $K^*$, $K_1$, and 
higher-mass Regge-poles. The dominance of unnatural parity exchange 
($K$- and $K_1$-exchanges) leads to a photon beam asymmetry equal to 
$-1$ while natural parity exchange ($K^*$-exchange) leads to a photon beam 
asymmetry equal to $+1$ at the limit of $t = 0$ and 
$s \rightarrow \infty$~\cite{Stichel,Guidal}.   
Therefore, measurements of the photon beam asymmetries will provide 
information relevant to $t$-channel meson exchange. 

In the present paper, we report photon beam asymmetries 
and the differential cross sections for the 
$\gamma p \rightarrow K^+ \Lambda$ and 
$\gamma p \rightarrow K^+ \Sigma^0$ reactions measured 
at $E_{\gamma}$=1.5$-$2.4 GeV at the SPring-8/LEPS facility. 
The photon beam asymmetries were obtained by using linearly polarized 
photons. The photon beam asymmetries, briefly reported in an earlier 
letter~\cite{KPRL}, are the first data in the nucleon resonance region.
In contrast to the CLAS detector, the LEPS spectrometer at SPring-8 covers 
forward scattering angles. The data presented here is thus complementary 
to the CLAS data set. 
Since there remains a significant discrepancy between the CLAS and 
SAPHIR cross section data especially at forward angles~\cite{CLAS}, 
new cross section data are important for solving this discrepancy.  

\section{EXPERIMENT}

\subsection{LEPS beam line at SPring-8 facility}

\begin{figure}[b]
\includegraphics[height=14cm]{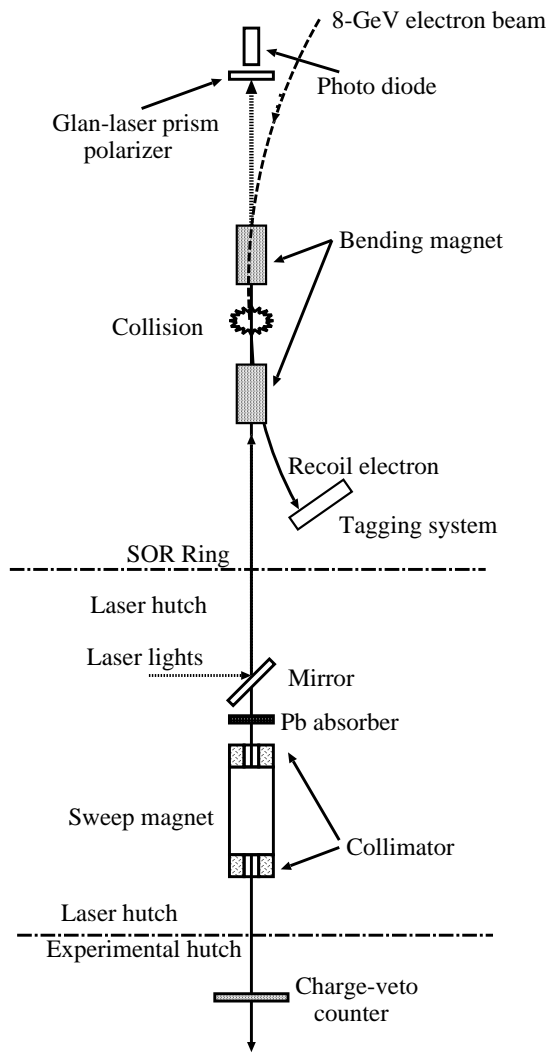}
\caption{\label{LEPSfacility}Schematic view of the LEPS facility at 
SPring-8.}
\end{figure}
The experiment was carried out at the Laser-Electron-Photon beam line 
(LEPS) at the storage ring of the Super Photon ring 8-GeV facility (SPring-8). 
At the LEPS beam line, a multi-GeV photon beam was produced by 
backward-Compton scattering of laser light 
from the circulating 8-GeV electrons. 

A schematic view of the LEPS beam line is shown in Fig.~\ref{LEPSfacility}. 
The direction and polarization of the laser was tuned using 
mirrors and a half wave-length plate. The laser light was reflected into 
the beam line and injected in a 7.8 m straight section of the 
storage ring. The backward-Compton process occurs when laser light collides 
with the 8-GeV electrons. 
The photons are scattered backward toward the experimental 
hutch where a target and a magnetic spectrometer are located.   

In the present experiment, an Ar-ion laser was used. Laser light with 
wave lengths ranging from 333.6 nm to 363.8 nm was obtained from the 
ultra-violet multi-line mode of operation.  Linearly-polarized laser 
light produces linearly-polarized GeV energy photons. 
The polarization of the laser light was measured using a Glan-laser prism 
polarimeter and a photo diode placed at the end of the straight section. 
The degree of polarization of the laser light was typically 98\%. 

\begin{figure}
\includegraphics[height=5.3cm]{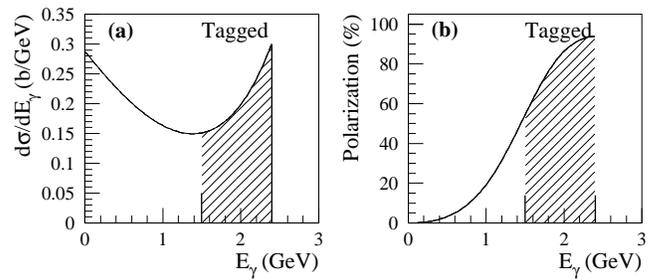}
\caption{\label{BCS}(a) Intensity distribution and (b) linear polarization 
of the photon beam as a function of photon energy ($E_\gamma$), calculated for 
the 8-GeV incident electron beam and laser light with a wavelength of 351 nm. 
The polarization of the laser is assumed to be 100\%. The shaded areas 
correspond to the photon energy region tagged by recoil electrons.}
\end{figure}
Figure~\ref{BCS}(a) shows the intensity distribution of photons produced 
by the backward-Compton process as a function of the photon 
energy~\cite{BCSEvsP}. 
The maximum energy of the photon beam was 2.4 GeV. 
The lowest energy of the tagged photon beam (see below) was 1.5 GeV. 
Figure~\ref{BCS}(b) shows the degree of the linear polarization of the photon 
beam as a function of photon energy~\cite{BCSEvsP}. 
The polarization of the photon beam was 95\% at the maximum 
energy, 2.4 GeV. The polarization drops as the photon energy decreases.  
However it is still as high as 55\% at 1.5 GeV. 
The typical photon intensity, integrated from 1.5 GeV to 2.4 GeV, was 
5 $\times 10^5$/s with a laser power of 5 W. 
The size of the photon beam at the target position which is about 
70 m from the collision point in the storage ring was $\sigma_x$ = 5 mm 
in the horizontal direction and $\sigma_y$ = 3 mm in the vertical direction. 

The energy of the photon beam was determined by measuring the energy of 
the recoil electron from Compton scattering with a tagging counter placed 
at the exit of the bending magnet next to the straight section (see 
Fig.~\ref{LEPSfacility}).  
\begin{figure}
\includegraphics[height=5cm]{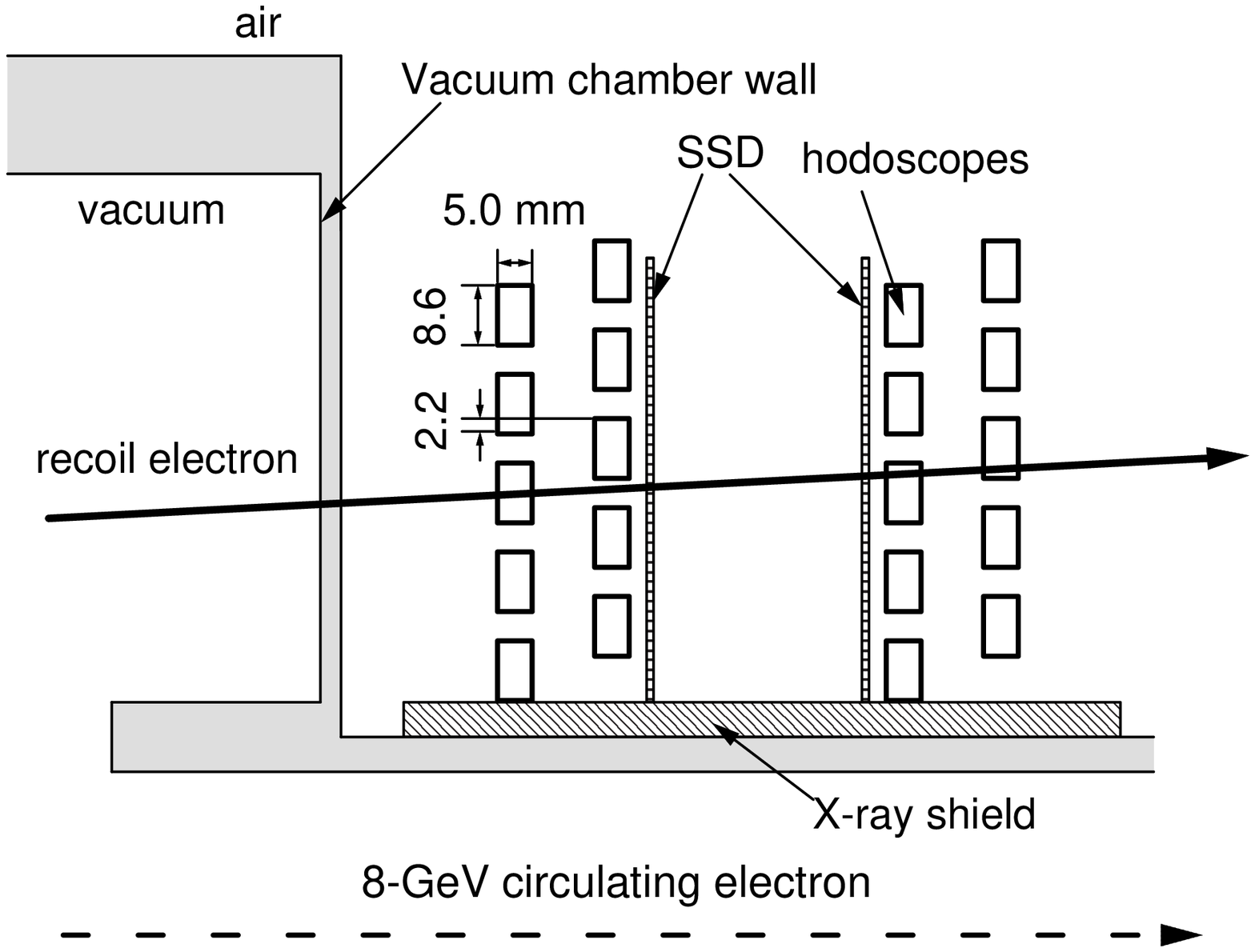}
\caption{\label{tagger}Schematic top view of the tagging counter.}
\end{figure}
Figure~\ref{tagger} shows the tagging counter which consists 
of two layers of scintillator hodoscopes and two layers of silicon strip 
detectors (SSD). 
The size of the scintillator was 10 mm high, 8.6 mm wide, and 5 mm thick. 
The hodoscope consisted of 10 plastic scintillation counters 
stacked with an overlap of 2.2 mm. 
The trigger required at least one hit in each layer of the hodoscope. 
The hodoscopes were used to reject accidental events. 
After finding a track candidate with two layers of the hodoscope,  
an associated hit at the SSD was found and the precise hit position of 
a recoil electron was measured by the SSD layers with a strip pitch of 0.1 mm. 
The photon energy coverage of the tagging counter was from 1.5 to 2.4 GeV.  
The photon energy resolution was 15 MeV in root mean square mainly stemming 
from the energy and angular spread of the 8-GeV circulating electrons and 
the uncertainty of the photon-electron interaction point.  

An aluminum coated silicon mirror of 0.85 mm thickness reflected the laser 
light, and a lead sheet of 2 mm thickness was also placed in the beam line. 
Its purpose was to absorb X-rays, thereby protecting the detectors from 
radiation damage. 
A part of the photon beam converted to charged particles in these materials.  
The charged particles were removed by a sweep magnet which consisted of 
a permanent magnet with a gap of 4.4 cm and a 100 cm length.  
The strength of the magnetic field was 0.6 T. 
Lead beam collimators were placed upstream and downstream of the sweep magnet. 
The upstream and downstream collimators had holes with a diameter of 20 mm 
and 25 mm, respectively. 
The size of those holes was much larger than the spread of the photon beam. 
Charged particles produced by the exit windows and the residual gas in the 
vacuum pipe between the sweep magnet and the target were eliminated 
using a charge-veto plastic scintillation counter placed just before 
the target.  

\subsection{LEPS spectrometer}

The target was liquid hydrogen in a cell with a trapezoid shape made of copper. 
The length of the target cell was 5 cm in the beam direction and 
the inner volume was 110 cm$^3$. 
The entrance and exit windows of the target cell were made of an Aramid foil 
of 0.05 mm thickness. The exit aperture was circular in shape with a 35 mm 
diameter. 

\begin{figure}
\includegraphics[height=7.5cm]{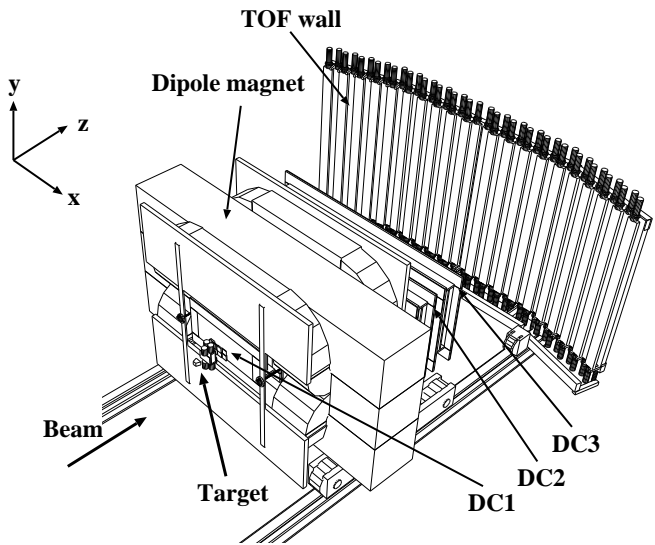}
\caption{\label{lepsdetector}The LEPS spectrometer.}
\end{figure}

\begin{figure}
\includegraphics[height=5cm]{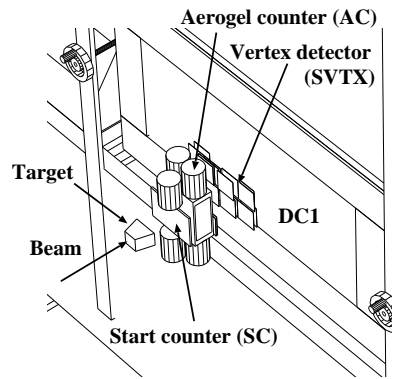}
\caption{\label{lepsdetector2}Detectors upstream of the dipole magnet. }
\end{figure}

The LEPS spectrometer, shown in Fig.~\ref{lepsdetector}, was designed 
to detect charged hadrons produced at forward angles. 
Figure~\ref{lepsdetector2} shows the setup in front of the dipole magnet 
in more detail. 
The spectrometer consisted of a start counter (SC), a silica-aerogel 
\v{C}erenkov counter (AC), a silicon vertex detector (SVTX), a dipole 
magnet, three multi-wire drift chambers (DC1, DC2 and DC3), and a 
time-of-flight (TOF) wall. 
 The field strength of the dipole magnet was 0.7 T at its center. 
The magnet aperture was 55 cm high and 135 cm wide.  The pole length 
was 60 cm along the beam direction.
The angular coverage of the spectrometer was about $\pm 0.4$ rad and 
$\pm 0.2$ rad in the horizontal and vertical directions, respectively. 

\begin{figure}
\includegraphics[height=5cm]{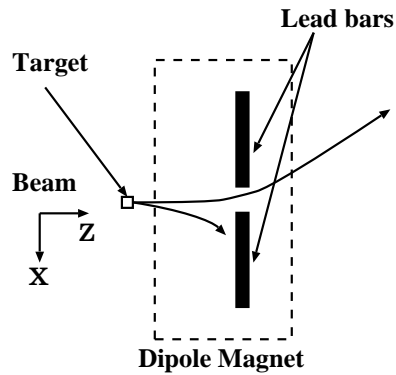}
\caption{\label{Pb-bar}Schematic top view of the lead bars. }
\end{figure}

Charged particles produced at the target hit the SC which is 
 5 mm thick and is located just behind the target. 
The signal from the SC determined the trigger timing for the data 
acquisition system. 
The main background in the measurement of photon induced hadronic reactions 
is $e^+e^-$ events from pair production. 
The AC had a 60 mm silica-aerogel radiator with a refractive index of 1.03 
and was used to reject $e^+e^-$ events at the trigger level. 
Gortex sheets were used as a reflector for the light collection. 
The rejection efficiency for $e^+e^-$ events was higher than 99.9\%. 
The magnetic field of the dipole magnet bent the trajectories of low 
momentum positrons and electrons.
To prevent radiation hazards, positrons and electrons with 
momenta below 1 GeV/c were blocked by two lead bars in the dipole magnet.  
Figure~\ref{Pb-bar} shows the top view of the lead bars. 
The lead bars with 4 cm height were located in the horizontal plane 
of the dipole magnet.   
The area covered by the lead bars corresponded to about 6\% of the aperture of 
the dipole magnet. The positrons and electrons with momenta above 
1 GeV/c went through the 15.5 cm gap between the two lead bars and 
into the beam dump positioned behind the TOF wall. 

The silicon vertex detector (SVTX) and three drift chambers 
were used as tracking devices.  SVTX was located behind the AC. 
By using the silicon detectors, precise vertex positions were obtained. 
The SVTX consisted of two layers of silicon-strip detectors with a 0.12 mm 
pitch. One layer was used to measure the position in horizontal direction 
and the other in the vertical direction. 
SVTX had a hole of 10 mm $\times$ 10 mm for the beam. 
One of drift chambers (DC1) was positioned upstream of the magnet 
and had 6 layers (three vertical layers, two layers at $+45^\circ$ and 
one layer at $-45^\circ$). 
The active area of DC1 was 30 cm high $\times$ 60 cm wide. 
The other two chambers (DC2 and DC3) were positioned at the downstream 
of the magnet and had 5 layers: 
two vertical layers, two layers at $+30^\circ$ and 
one layer at $-30^\circ$. The active area was the same for DC2 and DC3, at 
80 cm high $\times$ 200 cm wide. 
The gas mixture used to operate the drift chambers was 70\% argon and 30\%
isobutane. 
The position resolution of the drift chambers was 200 $\mu$m (RMS). 
The average efficiency of each layer was about 99\%. 

The TOF wall located at the downstream of the DC3 was an array of 
40 plastic scintillators. 
The dimension of each scintillation counter was 12 cm wide, 4 cm thick and 
200 cm high. The counters were placed with an overlap of 1 cm. 
Photo-Multiplier (PM) tubes were attached to the top and bottom of the 
counter. The typical time resolution of the TOF counters was 120 ps. 

The event trigger was made by 
signals from the tagging hodoscopes, the SC and the TOF wall.
Signals from the charge-veto counter and the AC were used as vetos. 
The typical trigger rate was 20 Hz. The dead time of the data 
acquisition system was about 3\%.  

Half of the data were taken with vertically polarized photons 
and the other half with horizontally polarized photons. 
The polarization was changed about every 6 hours to reduce  
systematic errors in the measurement of the photon beam asymmetries. 
The laser polarization was also measured about every 6 hours.   
The data were accumulated with 2.1 $\times 10^{12}$ photons 
at the target in total. 

\section{DATA ANALYSIS}

\subsection{Event reconstruction}

Momenta of charged particles were determined using information from the 
vertex detector and the three drift chambers. 
In the first stage of the tracking process, straight-line tracks 
were defined separately in the upstream devices (SVTX and DC1) 
and in the downstream devices (DC2 and DC3/TOF) to determine the hit position 
from the drift distance and the wire address.  
After straight-line tracking, track candidates were listed from 
all possible combinations of the upstream and downstream tracks. 
The overall track fitting was performed for all the track candidates using 
the full information of hits.   
The Kalman filter technique was employed for the track fitting 
taking into account the effect of the multiple scattering~\cite{Kalman}. 
The trajectory of a charged particle in the inhomogeneous magnetic field 
of the LEPS magnet was calculated by the Runge-Kutta method. 
Tracks fitted within a 98\% confidence level were accepted for further analysis. 

The vertical position in the TOF counter was calculated from the time 
difference of the TDC signals between the two PM tubes attached to the top and 
bottom of a counter. The resolution in vertical hit position was 
18 mm ($\sigma$). The vertical hit position was used to find the correct 
combination between a track and a TOF hit. 
The stop signal for the time-of-flight measurement was provided from signals 
of the TOF counters. 
The start signal was provided by the RF signal from the 8-GeV electron storage 
ring where electrons were bunched at 2 ns intervals with a width of 12 ps 
(RMS). 
The typical flight path of charged particles was 4.2 m. 
Since the electronics used for recording the RF timing was 
not working for a part of the experiment, the start counter (SC) was 
used to provide a start timing instead of the RF signal for two-thirds of 
the data. 
The time resolution of the SC was 150 ps, and the time-of-flight 
resolution was 180 ps. 

\begin{figure}
\includegraphics[height=7cm]{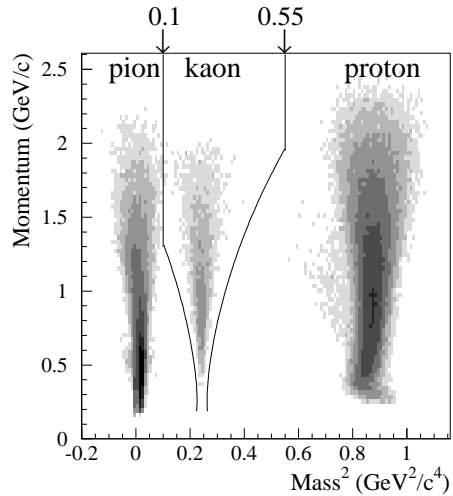}
\caption{\label{MassCut}Two-dimensional scatter plot of momentum vs. 
mass squared. 
Cut boundaries (3$\sigma$) of the PID selections are displayed.}
\end{figure}

The particle mass was calculated using the momentum, the path length and the  
time-of-flight. 
Figure~\ref{MassCut} shows the momentum versus the square of mass.
The mass resolution depends on the momentum, as one can see in the plot. 
The resolution of kaon mass was 30 (105) MeV/c$^2$ at the 1 GeV/c (2 GeV/c)
momentum. 

\subsection{Event selections}

To select events from the $\gamma p \rightarrow K^+ \Lambda$ and 
$\gamma p \rightarrow K^+ \Sigma^0$ reactions, we required the following 
cut conditions: 
(1) select $K^+$ particles, 
(2) remove decay-in-flight kaons, 
(3) reject accidental $e^+e^-$ events, 
(4) select reaction vertices in the target, 
(5) select a recoil electron by Compton scattering and to reject shower 
    and accidental hits in the tagging counter and,  
(6) select either $\Lambda$ or $\Sigma^0$ production.  

The $K^+$ particles were selected using the known mass and charge. 
The curves in Fig.~\ref{MassCut} indicate the 3$\sigma$ boundary of the 
momentum-dependent mass resolution for kaons. The 3$\sigma$ cut was used 
to select kaons with an additional condition of 0.1 $<$ Mass$^2$ 
$<$ 0.55 in the overlap region with positive pions and protons at 
high momenta. 
Events were purified by selecting events within the 98\% confidence level 
for track fitting.  This selection  
rejected mainly decay-in-flight kaons. The position difference  
between a track and a TOF hit was also used to reject decay-in-flight 
kaons. 
Although most of the accidental $e^+e^-$ events were rejected in  
the $K^+$ particle selection, there still remained a contamination of 
accidental $e^+e^-$ events. Those events were rejected by removing 
events with particles emitted in the median plane and passing through 
the gap in the lead bars.  

\begin{figure}
\includegraphics[height=7cm]{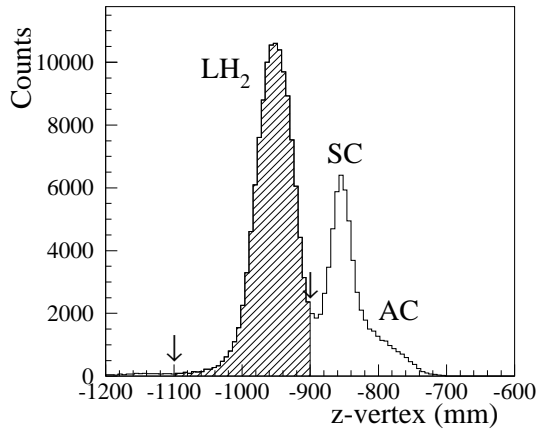}
\caption{\label{Vertex}Distribution of vertex position along the beam 
direction ($z$-vertex). 
The liquid hydrogen target (LH$_2$) and the start
counter (SC) positions are easily seen. The shoulder around $-$800 mm 
corresponded to the events from the aerogel counter (AC). 
Cut positions are indicated by the shaded area.} 
\end{figure}

Events produced at the liquid hydrogen target (LH$_2$) were selected 
by their calculated vertex position along the photon beam direction 
($z$-vertex).  
Figure~\ref{Vertex} shows the $z$-vertex distribution.  
The vertex point was defined as the point of the closest approach between 
a track and the beam axis. 
The photon beam had a small but finite size and we had no information on 
the position of a photon at the target on an event-by-event basis. 
The cut condition to select events produced at the target was 
$-$1100 mm $<$ $z$-vertex $<$ $-$900 mm.  
The downstream cut point was tightened to reduce contamination 
events from the SC. 

The number of electron tracks reconstructed in the tagging counter was 
required to be 1. Electro-magnetic shower events or accidental events 
could make a trigger, but these background events were rejected by 
requiring one track. 
After selecting $K^+$ particles and calculating the photon energy from the 
hit position of a recoil electron in the tagging counter, the missing mass 
of the $\gamma p \rightarrow K^{+} X$ reaction was calculated to identify 
$\Lambda$ and $\Sigma^0$ particles. 
\begin{figure}
\includegraphics[height=7cm]{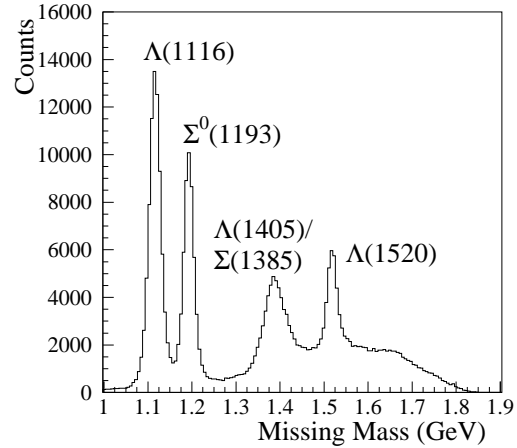}
\caption{\label{MissingMass}Missing mass of the 
$\gamma p \rightarrow K^{+} X$ reaction. 
The peaks correspond to the $\Lambda$(1116),
 $\Sigma^0$(1193) and hyperon resonances.}
\end{figure}
Figure~\ref{MissingMass} shows the missing mass   
spectrum of the $K^+$ photoproduction. Peaks corresponding to 
$\Lambda$ and $\Sigma^0$ were observed. 
\begin{figure}
\includegraphics[height=7cm]{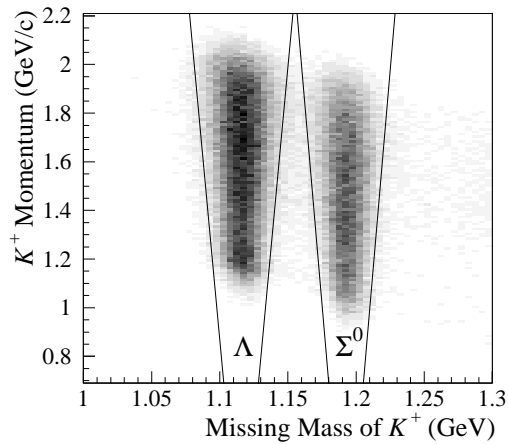}
\caption{\label{MMpp}Two-dimensional scatter plot of momentum vs missing mass 
of $K^+$. The solid lines represent missing mass cuts corresponding to 
2$\sigma$ mass deviations from mean values.} 
\end{figure}
The missing mass resolution depends on the $K^+$ momentum as shown 
in Fig.~\ref{MMpp}. 
The mass resolution of $\Lambda$ and $\Sigma^0$ particles 
are $\sigma_{mass}$ = 17 MeV/c$^2$ at a 2.0 GeV/c $K^+$ momentum and 
$\sigma_{mass}$ = 11 MeV/c$^2$ at a 1.2 GeV/c $K^+$ momentum. 
The 2$\sigma_{mass}$ boundaries 
were used to select $\Lambda$ and $\Sigma^0$ particles. 

\begin{table}
\caption{\label{tab:counts0}The number of events after selection
cuts for the beam polarization asymmetry analysis. }
\begin{ruledtabular}
\begin{tabular}{cc}
Selection cuts & Events \\  \hline
triggered events    & 1.78 $\times$ 10$^8$ \\
single track reconstructed& 4.23 $\times$ 10$^7$ \\
$K^+$ selection  & 1.67 $\times$ 10$^6$ \\
$K^+$ decay-in-flight rejection & 1.52 $\times$ 10$^6$ \\
$e^+e^-$ rejection   & 9.70 $\times$ 10$^5$ \\
LH$_2$ target selection by z-vertex & 4.52 $\times$ 10$^5$ \\
one recoil electron in tagger & 3.39 $\times$ 10$^5$ \\
$\Lambda/\Sigma^0$ particles & 7.25/4.89 $\times$ 10$^4$ \\
\end{tabular}
\end{ruledtabular}
\end{table}

Table~\ref{tab:counts0} shows the number of events surviving 
after the selection cuts. 
From the total set of 1.8 $\times$ 10$^8$ triggered events, 
7.3 $\times$ 10$^4$ and 4.9 $\times$ 10$^4$ 
events of the $K^+ \Lambda$ and 
$K^+ \Sigma^0$ reactions 
 satisfied all the cut conditions given above. 
 
\section{RESULTS}

\subsection{Photon beam asymmetries}

By using both vertically and horizontally polarized photon beams, 
two sets of data were accumulated to measure the photon beam asymmetries
~\cite{KPRL}. The relation between production yields in the two sets of 
data and the photon beam asymmetry, $\Sigma$, is given as follows:    
\begin{eqnarray}
\hspace{0.1cm}P_\gamma\Sigma \cos2\Phi=\frac{n \cdot N_{v}-N_{h}}{n \cdot 
N_{v}+N_{h}},
\label{eq:asymmetry}
\end{eqnarray}
where $N_{v}$ and $N_{h}$ are the $K^+$ photoproduction yields with the 
vertically and horizontally polarized photons, respectively, and 
$n$ is the normalization factor for $N_{v}$, determined by using 
the numbers of horizontally polarized photons, $n_{h}$, and vertically 
polarized photons, $n_{v}$, at the target as $n = n_{h}/n_{v}$. 
The value of $n$ is $0.923$ in the present experimental data. 
$\Phi$ is the $K^+$ azimuthal angle defined by the angle between 
the reaction plane and the horizontal plane, and $P_\gamma$ is 
the polarization degree of the photon beam. 
\begin{figure}  
\includegraphics[height=9cm]{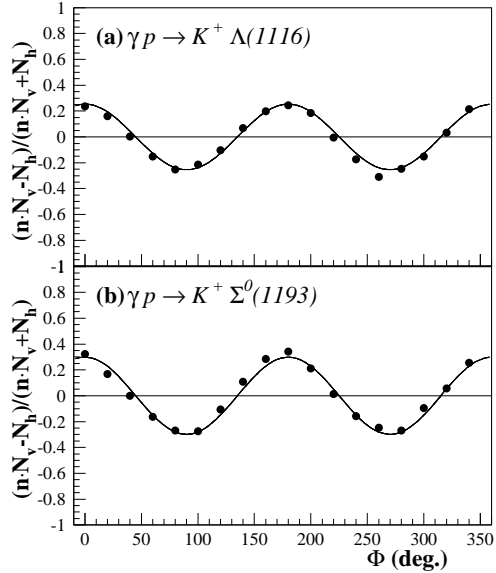}
\caption{\label{phiall}Azimuthal angle $\Phi$ dependence of the ratio 
$(n \cdot N_{v}-N_{h})/(n \cdot N_{v}+N_{h})$ in Eq.($\ref{eq:asymmetry}$) 
for (a) the $\gamma p \rightarrow K^+ \Lambda$ reaction and 
(b) $\gamma p \rightarrow K^+ \Sigma^0$ reaction integrating over all 
events. The solid lines are the result 
of fitting using a function of cos2$\Phi$. }
\end{figure}
The $\Phi$ dependence of the ratio 
$(n \cdot N_{v}-N_{h})/(n \cdot N_{v}+N_{h})$ 
was fitted with the function $\cos2\Phi$ as shown in
Fig.~\ref{phiall} and the amplitude $P_{\gamma}\Sigma$ was obtained.  
After $P_{\gamma}$ was calculated, 
using the photon energy $E_\gamma$ and the laser polarization 
shown in Fig.~\ref{BCS}(b), the photon beam asymmetry $\Sigma$ was obtained. 

The contamination of positive pions and protons into the $K^+$ mass region 
was estimated by extrapolating the Gaussian shaped mass distributions of 
positive pions and protons into the $K^+$ region. 
The contamination rates increased with $K^+$ momentum 
since the mass resolution was poor at the higher momenta. 
The contamination rates 
of positive pions and protons for the $K^+\Lambda$ 
at the highest $K^+$ momentum region were 2.0\% and 2.5\%, 
respectively. The contamination rates of positive pions and 
protons for the $K^+\Sigma^0$ were 3.5\% and 5.0\% 
at the highest $K^+$ momentum region, respectively. Although the 
contamination rates were small, their contribution caused a non-negligible 
shift of the measured photon beam asymmetry. 
The measured photon beam asymmetry is written as 
$\Sigma^{meas}_{K^+} = (1 - C_{bg}) \Sigma_{K^+} + C_{bg} \Sigma_{bg}$. 
The photon beam asymmetry of the $K^+$ events, $\Sigma_{K^+}$, was 
obtained by determining the contamination rate, $C_{bg}$, and the photon 
beam asymmetry of the background events, $\Sigma_{bg}$. 
The maximum correction for the contamination of positive pions and protons  
was $\delta\Sigma$ = 0.026 at W=2.28 GeV ($E_\gamma$=2.3 GeV). 
Another correction was made for background from the SC, 
estimated as a function of the $K^+$ scattering angle. 
The contamination rate was lower than 1\% at cos$\Theta_{cm}<$0.9.  
It was 1.5\% and 6.5\% for the $K^+\Lambda$ events 
at cos$\Theta_{cm}$=0.925 and 0.975, respectively.  
For the $K^+\Sigma^0$ events, it was 2.5\% and 11\%   
at cos$\Theta_{cm}$=0.925 and 0.975, respectively.  
The contamination rate for the $K^+\Sigma^0$ reaction was about 
 2 times higher than that for the $K^+\Lambda$ reaction   
because $\Sigma^-$(1197) events,  
from the $\gamma n \rightarrow K^+ \Sigma^-$ reaction 
in carbon from the SC, were also present in 
the missing mass selection of the $\Sigma^0$ events. 

\begin{figure}
\includegraphics[height=13cm]{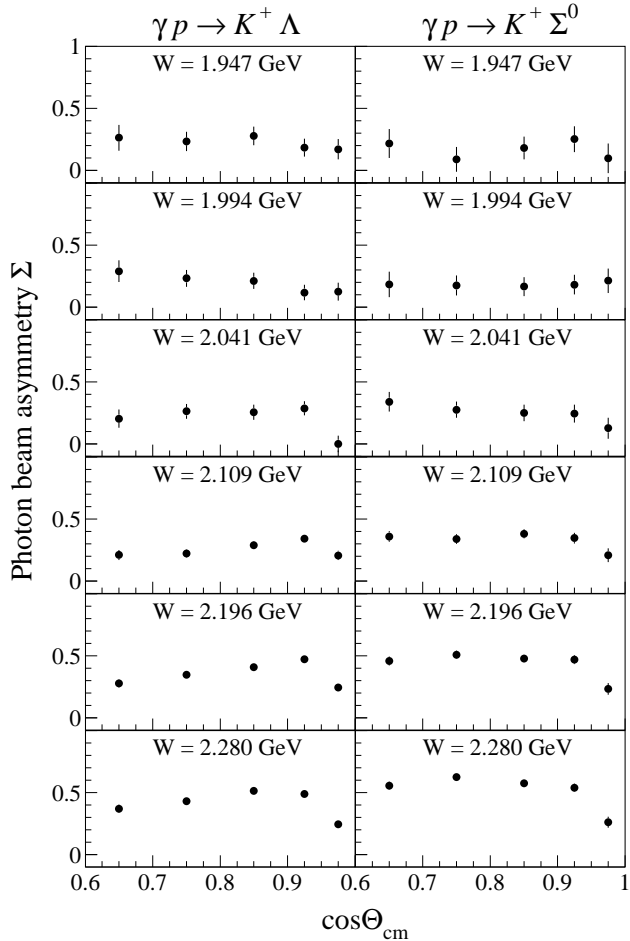}
\caption{\label{Asymmetry}Photon beam asymmetries for the
 $\gamma p \rightarrow K^+ \Lambda$ (left) and 
$\gamma p \rightarrow K^+ \Sigma^0$ (right) reactions as a function of 
cos$\Theta_{cm}$. }
\end{figure}

Figure~\ref{Asymmetry} shows the experimental results of the photon beam 
asymmetries as a function of cos$\Theta_{cm}$ for the $K^+\Lambda$ and 
$K^+\Sigma^0$ reactions 
in the photon energy range from 1.5 to 2.4 GeV. The errors are statistical 
only. Systematic uncertainties resulted from: 
(1) the photon-yield normalization factor $n$, 
(2) the polarization degree, and 
(3) the polarization direction of the photon beam. 
To investigate the accuracy of the normalization factor 
obtained from the counting rate of the tagging counter, 
a fitting was performed with a function of p$_1\cdot$cos2$\Phi$ 
+ p$_2$ including an additional offset parameter of p$_2$.  
As the result of the fitting, the systematic error for the photon 
asymmetries was estimated to be $\delta\Sigma$ = $-$0.02 $\sim$ $+$0.06. 
The measurement error of the laser polarization
 was estimated to be $\pm 1.5\%$ ($|\delta\Sigma| <$ 0.01). 
The polarization, $P_\gamma$, was calculated by assuming that the wavelength 
of laser light was 351 nm (which was the most dominant component). 
The systematic error of $P_\gamma$ due to the uncertainty of the wavelength 
was smaller than 0.1\%.
The direction of the polarization deviated from 0$^\circ$ and 90$^\circ$
 in the cases of horizontal and vertical polarization, respectively. 
The deviation was 4$^\circ$ at a maximum.  
The systematic error arising from this deviation was estimated to be smaller 
than 0.4$\%$ ($|\delta\Sigma| <$ 0.001). 

The signs of the photon beam asymmetries for both of the 
$K^+\Lambda$ and 
$K^+\Sigma^0$ reactions were found to be positive 
in the measured kinematical region. 
The positive sign means that $K^+$ particles are emitted preferentially 
in the orthogonal direction to the photon polarization.
The photon beam asymmetry increases with increasing photon energy in both 
reactions.  
The photon beam asymmetries for the $K^+\Lambda$ 
reaction slightly decrease at backward angles above W=2.0 GeV,  
while showing a flat angular distribution below W=2.0 GeV.  
The photon beam asymmetries for the 
$K^+\Sigma^0$ reaction show a flat 
angular distribution in all energy regions.  

The ESRF/GRAAL collaboration reported the measurements of 
photon beam asymmetries for the $K^+\Lambda$ 
reaction at total energies up to 1.87 GeV ($E_\gamma$=1.4 GeV)~\cite{GRAAL}. 
Their preliminary experimental analysis resulted in a positive sign and 
a flat angular distribution over all angles at $W$=1.87 GeV. 
The GRAAL data show a good connection to the LEPS data 
around $W$=1.92 GeV ($E_\gamma$=1.5 GeV). 

\subsection{Differential cross sections}

When the SC was used to provide the time-of-flight start timing, 
multi-hit events from a fast proton (from hyperon decay) and a $K^+$ 
particle caused a deterioration of the time resolution.  This led to 
an efficiency loss in the particle identification. In the analysis of cross 
sections, we used only the part of the data with a good quality RF 
signal.

The data set used in this analysis were accumulated with the same 
number of photons for the horizontal and vertical polarizations ($n_v=n_h$). 
Since the $K^+\Lambda$ and $K^+\Sigma^0$ reactions have finite values of the 
photon beam asymmetry, 
the production yield, $N_{K^+}$, was the average of yields obtained with 
vertically and horizontally polarized photon beams.  
Differential cross sections were calculated as follows:
\begin{eqnarray}
\frac{d\sigma}{d\cos\Theta_{cm}}
= \frac{(N_{K^+} - N_{BG})/a}{N_{B}N_{T}\Delta\cos\Theta_{cm}},  
\label{eq:cross_section}
\end{eqnarray}
where $a$ is for the correction of the efficiency of the $K^+$ selection 
including the detector acceptance,  
and $N_{BG}$ is a correction for the contamination of background. 
$N_{T}$ is the number of protons inside the target cell and $N_{B}$ is  
the number of photons at the target. 
In this analysis, $\Delta\cos\Theta_{cm}$ was 0.1. 

When the target cell was filled with liquid hydrogen, the typical pressure 
and temperature were 1.05 atm and 20.0 K, respectively and $N_T$ was 2.37 
$\times 10^{23}$ protons/cm$^2$. The photon number, $N_B$ was obtained 
by counting the number of hits in hodoscopes of the tagging counter. 
The efficiency of the hodoscopes and the acceptance of finding 
a track of a recoil electron were taken into account. 
The discriminator dead time in electronics, due to a high count rate in the 
hodoscopes, was corrected 
to obtain the number of photons.  The photon beam transmission 
from the collision point in the storage 
ring to the target position was measured with a PbWO$_4$ crystal  
calorimeter. 
The result of the transmission measurement was consistent with an estimated 
value 
from the photon beam loss by material in the LEPS beam line. 

The conditions to select the 
$K^+\Lambda$ and $K^+\Sigma^0$ reaction events are the same as 
those used in the analysis of the photon beam asymmetries. 
Table~\ref{tab:counts} shows the number of events which survive 
after the selection cuts. 
\begin{table}
\caption{\label{tab:counts}The number of events after the selection
cuts for the differential cross section analysis. }
\begin{ruledtabular}
\begin{tabular}{cc}
Selection cuts & Events \\ \hline
triggered events    & 3.68 $\times$ 10$^7$ \\
single track reconstructed& 7.68 $\times$ 10$^6$ \\
$K^+$ selection  & 2.80 $\times$ 10$^5$ \\
$K^+$ decay-in-flight     & 2.59 $\times$ 10$^5$ \\
$e^+e^-$ rejection   & 2.46 $\times$ 10$^5$ \\
LH$_2$ target selection by z-vertex & 1.22 $\times$ 10$^5$ \\
one recoil electron in Tagger & 9.11 $\times$ 10$^4$ \\
$\Lambda/\Sigma^0$ particles & 2.19/1.45 $\times$ 10$^4$ \\
\end{tabular}
\end{ruledtabular}
\end{table}
In total, 2.19 $\times$ 10$^4$ and 1.45 $\times$ 10$^4$ 
events for the $K^+\Lambda$ and 
$K^+\Sigma^0$ reactions, respectively, were used for the 
differential cross sections. 

The acceptance of $K^+$ particles was estimated by assuming a Gaussian 
shape of the mass distributions. The acceptance of the $K^+$ selection 
was 94\% at the highest momenta, as the cut position was set tighter 
there due to deteriorated mass resolution. 
The acceptance of selecting $K^+$ particles and hyperons   
was corrected by the factor of $a$ in Eq.(2).  
The amounts of contamination of positive pions and protons in the $K^+$ event 
samples were 1.7\% and 2.2\% for $K^+\Lambda$ and $K^+\Sigma^0$ 
production, respectively, in the highest $K^+$ momentum region. 
The contamination from the SC was the same as that in the photon asymmetry 
analysis. The contaminations of $\Sigma^0$ in the $K^+\Lambda$ sample and 
$\Lambda$ in the $K^+\Sigma^0$ sample were smaller than 1\%.  
The contamination from the target windows was estimated to be 4.2\%.  
These backgrounds were subtracted from the yield, $N_{K^+}$.  

The spectrometer acceptance, including the efficiency for track reconstruction, 
was estimated using a simulation tool based on the GEANT3 package. 
The acceptance, which depended on the photon energy and the ${K^+}$ scattering 
angle, was calculated for the $K^+\Lambda$ and $K^+\Sigma^0$ reactions. 
In Table~\ref{tab:mceff}, the acceptance is listed for the $K^+\Lambda$ 
reaction with selection cuts at each scattering angle, cos$\Theta_{cm}$=0.75,
 0.85 and 0.95, integrated over all photon energies. 
The acceptance for the $K^+\Sigma^0$ reaction is almost the same as 
that for the $K^+\Lambda$ reaction, due to only 
small difference of the $K^+$ momentum. 
\begin{table}
\caption{\label{tab:mceff}Acceptance rate with the selection cuts for
 the $\gamma p \rightarrow K^+ \Lambda$ reaction obtained by the simulation.}
\begin{ruledtabular}
\begin{tabular}{cccc}
&& cos$\Theta_{cm}$ range\\ 
Cuts  &0.9-1.0 & 0.8-0.9 & 0.7-0.8\\ \hline
Triggered events         & .645 & .399 & .258 \\
$K^+$ decay-in-flight rejection & .974 & .970 & .965 \\
$e^+e^-$ rejection       & .954 & .995 & .992 \\
LH$_2$ target selection by z-vertex  & .928 & .977 & .988 \\
Total                    & .556 & .376 & .244 \\
\end{tabular}
\end{ruledtabular}
\end{table}
The vertex resolution was poor at forward scattering angles, 
which caused a high rejection rate by the $z$-vertex cut at 
cos$\Theta_{cm}= 0.95$. 
About 56\%, 38\% and 24\% of the $K^+$ events were detected, and the mass was 
reconstructed, for cos$\Theta_{cm}$=0.95, 0.85 and 0.75, 
respectively. 

\begin{figure}
\includegraphics[height=9.cm]{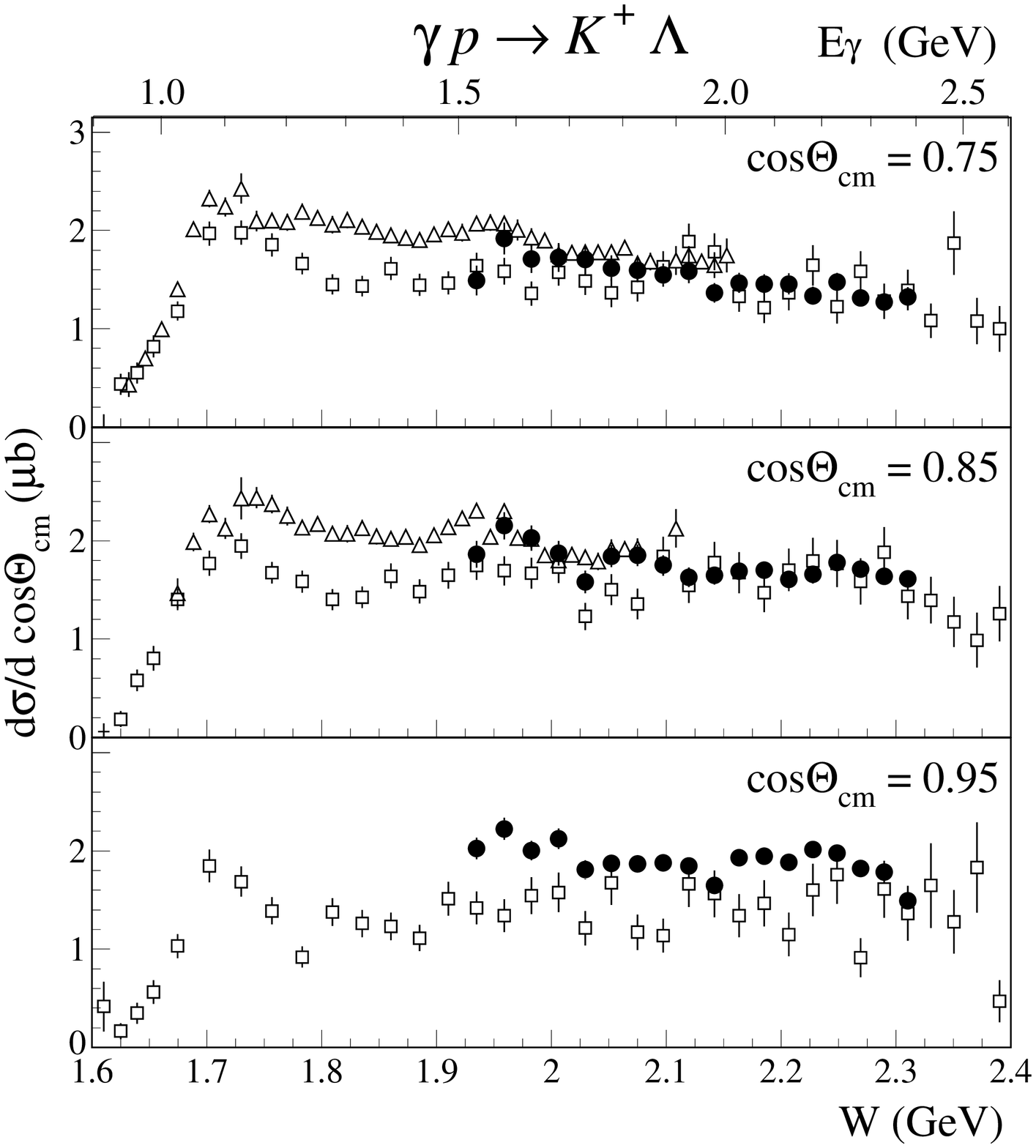}
\caption{\label{CrossLambda}Energy dependence of differential cross 
sections for the $\gamma p \rightarrow K^+ \Lambda$ reaction. 
The closed circles are the results of the present analysis. 
The open squares and triangles are the data measured by the
SAPHIR ~\cite{saphirk+} and the CLAS~\cite{CLAS}
collaborations, respectively. Errors are only due to the statistical one.}
\end{figure}
\begin{figure}
\includegraphics[height=9.cm]{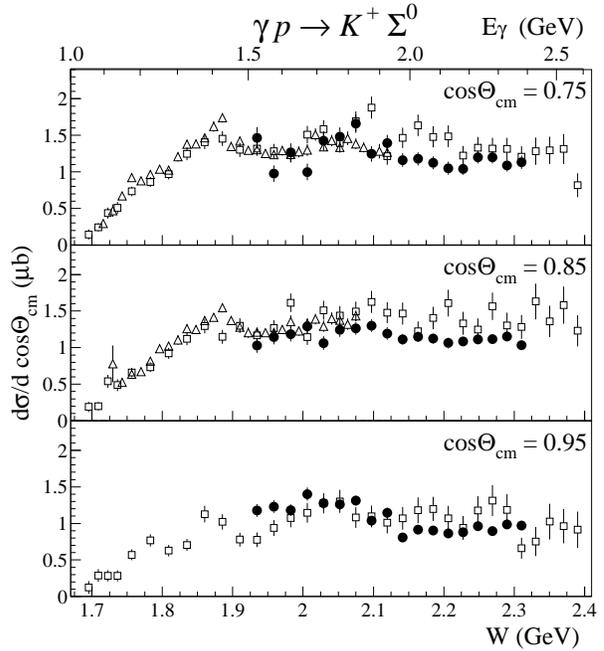}
\caption{\label{CrossSigma}Energy dependence of differential cross 
sections for the $\gamma p \rightarrow K^+ \Sigma^0$ reaction. 
The closed circles are the results of the present analysis. 
The open squares and triangles are the data measured by the
 SAPHIR ~\cite{saphirk+} and the CLAS~\cite{CLAS} collaborations, 
respectively. Errors are only due to the statistical one.}
\end{figure}

Figures~\ref{CrossLambda} and~\ref{CrossSigma} present results 
of differential cross sections (closed circles) for the $K^+\Lambda$ 
and $K^+\Sigma^0$ reactions. 
Differential cross section data for the $K^+\Lambda$ 
and $K^+\Sigma^0$ reactions were obtained 
with good statistics in the $W$ range from 1.92 to 2.32 GeV at 
cos$\Theta_{cm}$=0.95.
The systematic uncertainty of the target thickness, due to 
fluctuations of the temperature and pressure of the liquid hydrogen, 
is estimated to be 1.0\%. 
The systematic error of the photon number normalization was estimated 
using the number of protons detected by the LEPS spectrometer. 
The fluctuation of the photon number was within 1.2\%. 
The measurement error in the transmission of the photon beam was 3.0\%.  
The systematic uncertainty of the aerogel \v{C}erenkov counter (AC) due to 
accidental vetoes and $\delta$-ray reactions was measured to be 
lower than 1.6\%. 

The results from the SAPHIR Collaboration (triangle) 
and the CLAS Collaboration (square) are 
also plotted in Figs. 13 and 14 for comparison. 
Differential cross sections for the $K^+\Lambda$ 
reaction gradually decrease as the photon energy increases as shown in
Fig.~\ref{CrossLambda}. 
There is the large discrepancy between the CLAS and SAPHIR data at 
cos$\Theta_{cm}$=0.75 and 0.85, and this causes difficulty in 
simultaneous fitting for theoretical models.
It is important to solve these experimental discrepancies in order to 
obtain a conclusion for the existence of new resonances.
The LEPS data shows good agreement with the CLAS data within the
systematic uncertainty.
The LEPS data support the results of CLAS rather than the SAPHIR data.
A discrepancy between the LEPS and the SAPHIR data is seen at 
cos$\Theta_{cm}$=0.95. The LEPS cross section is 32\% larger than the 
SAPHIR cross section at $W<2.1$ GeV.
The difference is significant at $W\sim 1.95$ GeV.

Although a small bump structure is seen at $W\sim 1.96$ GeV in
the CLAS and SAPHIR data at cos$\Theta_{cm}$=0.75 and 0.85, 
the discrepancy is large. The LEPS data shows the bump structure at 
W$\sim$1.96 GeV and supports the results of the CLAS data.  
At cos$\Theta_{cm}$=0.95, the LEPS data still have the bump structure 
at W$\sim$1.96 GeV. No prominent structure is seen in the SAPHIR data.

Differential cross sections for the $K^+\Sigma^0$ reaction increase 
slightly at $W\sim 2.05$ GeV in both the LEPS and CLAS data. 
The cross sections for the $K^+\Sigma^0$ reaction are about 20\%
smaller than those for the $K^+\Lambda$ reaction.
The LEPS data are about 10\% smaller than the SAPHIR data
at $W>2.1$ GeV.  At cos$\Theta_{cm}$=0.95, the discrepancy between 
the LEPS and SAPHIR data is large around $W\sim 1.93$ GeV. 

\section{DISCUSSION}
\subsection{Photon beam asymmetries}

\begin{figure}
\includegraphics[height=12.5cm]{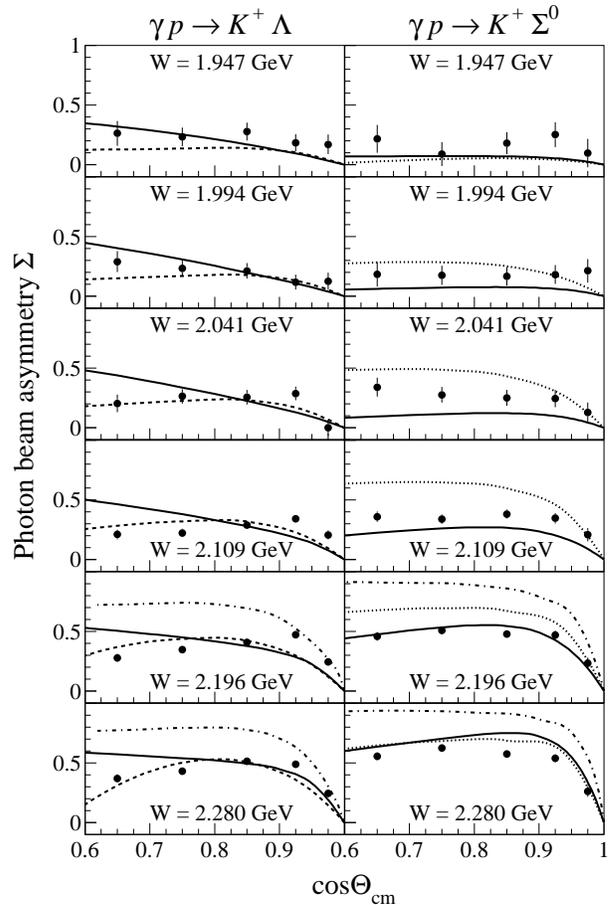}
\caption{\label{Asymmetry2}Photon beam asymmetries for the $\gamma p 
\rightarrow K^+ \Lambda$ (left) and $\gamma p \rightarrow K^+ \Sigma^0$ 
(right) reactions as a function of cos$\Theta_{cm}$. 
The dot-dashed curves are the results of the Regge model
with $K$  and $K^*$ exchanges by Guidal {\em et al.}~\cite{Guidalnew}. 
The dashed curves in the $K^+\Lambda$ reaction are the results of 
the Ghent isobar model by D.G. Ireland {\em et al.}~\cite{Janssennew}. 
The dotted curves in the $K^+\Sigma^0$ reaction are the results of 
the Ghent isobar model by T. Corthals {\em et al.}~\cite{corthalscomi}. 
The solid curves are the results of the mixing model of the Feynman 
diagram and the Regge model by Mart and Bennhold~\cite{bennholdcomi}.}
\end{figure}

\begin{table*}
\caption{\label{tab:model}List of the theoretical models. 
The $N^*$ resonances, $S_{11}(1650)$, $P_{11}(1710)$, $P_{13}(1720)$ and 
$D_{13}$(1900), are for the $K^+\Lambda$ and $K^+\Sigma^0$ reactions. 
The $\Delta^*$ resonances, $S_{31}$(1900) and $P_{31}$(1910), are for 
the $K^+\Sigma^0$ reaction only. 
In the Ghent model calculation, $\Lambda$(1800) 
and $\Lambda$(1810) resonances are for the $K^+\Lambda$ reaction, and 
$\Lambda$(1810) and $\Lambda$(1880) are  for the $K^+\Sigma^0$ reaction.}
\begin{ruledtabular}
\begin{tabular}{c|c|c|c|c}
Name & Model & t-channel & s-channel & u-channel  \\  \hline
Guidal~\cite{Guidalnew2} & Regge & $K, K^*$ & None & None \\
Ghent group~\cite{Janssennew,Janssennew2,corthalscomi}& Isobar & $K, K^*, K_1$ &  $S_{11}$, $P_{11}$, $P_{13}$, $D_{13}$, $S_{31}$, $P_{31}$& $\Lambda$(1800), $\Lambda$(1810),  $\Lambda$(1880)\\
Mart-Bennhold~\cite{bennholdcomi} & Isobar + Regge & $K, K^*, K_1$ & $S_{11}$, $P_{11}$, $P_{13}$, $D_{13}$, $S_{31}$, $P_{31}$ & None\\
\end{tabular}
\end{ruledtabular}
\end{table*}

The experimental data of photon beam asymmetries are compared with 
the results of theoretical calculations in Fig.~\ref{Asymmetry2}. 
The models used in the calculations are listed in Table~\ref{tab:model}. 
The dot-dashed curves are the results of the Regge model with the 
$K$ and $K^*$ exchanges by Guidal {\em et al.}~\cite{Guidalnew}. 
The dashed curves in the $K^+\Lambda$ reaction are the results of 
the Ghent isobar model by D.G. Ireland {\em et al.}~\cite{Janssennew}. 
The dotted curves in the $K^+\Sigma^0$ reaction are the results of 
the Ghent isobar model by T. Corthals {\em et al.}~\cite{corthalscomi}. 
The solid curves are the results of the mixing model of the Feynman diagram 
and the Regge model by Mart and Bennhold~\cite{bennholdcomi}.

The results of the Regge model calculation~\cite{Guidalnew2} are 
compared with the data at $W$=2.196 and 2.280 GeV, where the $t$-channel 
contribution is expected to become large. The photon beam asymmetry 
is a good means to study meson-exchange in the $t$-channel. 
The Regge model is valid only at forward angles and at high energies, 
and the $s$-channel contribution seems to be not negligible even at 
$W$=2.1$-$2.3 GeV.  
Although the Regge model calculation indicates a sharp rise at forward 
angles, which the data show, the model overestimates the data in both 
reactions and the discrepancy between the results of the model 
calculation and the experimental data becomes large at backward angles. 

The Ghent isobar-model calculation for the $K^+\Lambda$ reaction 
agrees with the LEPS data except for a sharp rise at forward angles. 
The model calculation for the $K^+\Sigma^0$ reaction mostly agrees 
with the data at $W$=2.28 GeV, but overestimates the data at $W<2.2$ GeV. 
the Mart and Bennhold model calculation for the $K^+\Lambda$ reaction 
mostly agrees with the data, but shows a discrepancy with the data 
at $\cos\Theta_{cm}<$0.75 and cannot reproduce the sharp rise at 
forward angles. The model calculation for the $K^+\Sigma^0$ reaction 
mostly agrees with the data. 

\subsection{Differential cross sections}

\begin{figure}
\includegraphics[height=9.cm]{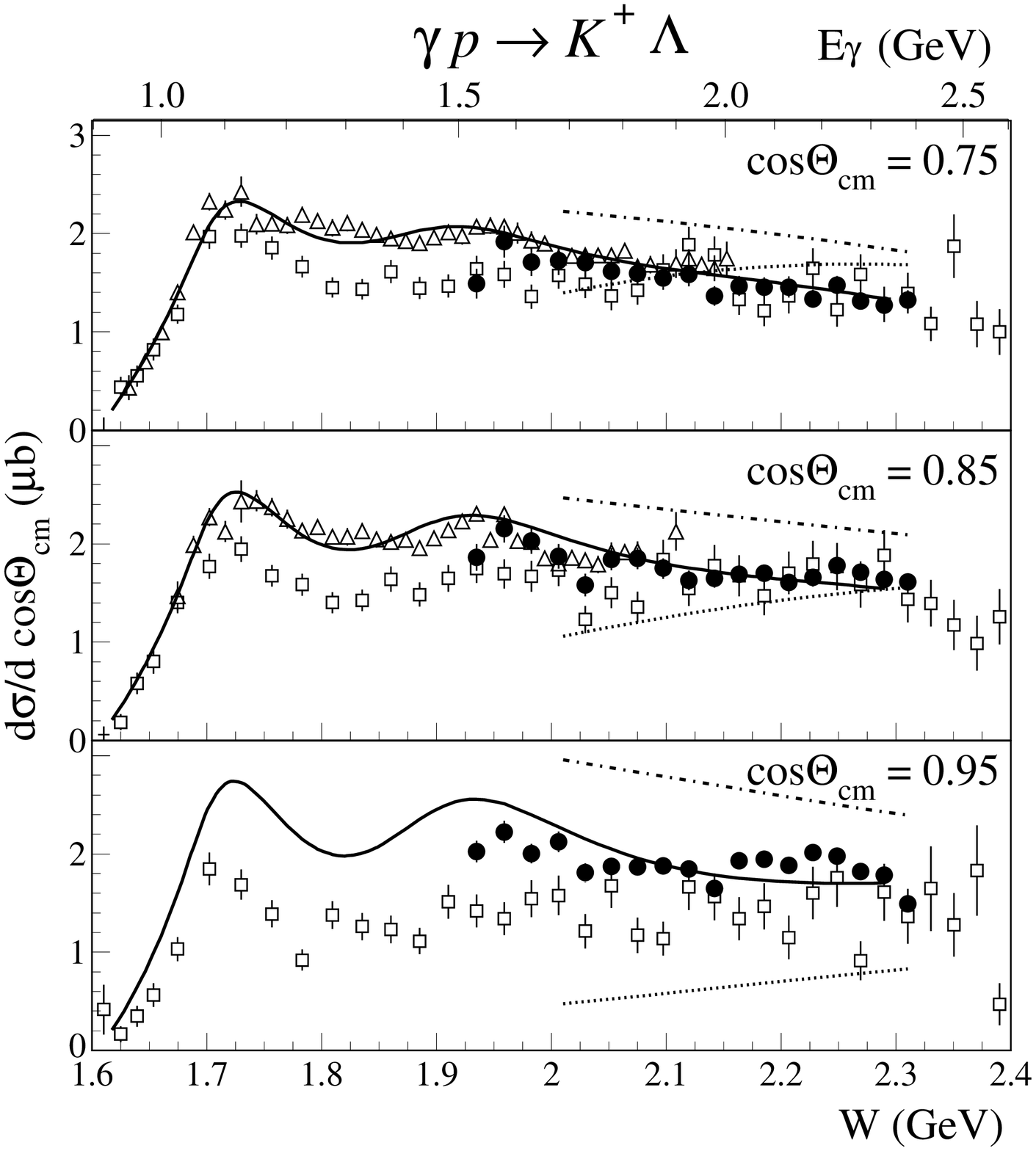}
\caption{\label{CrossLambda2}Energy dependence of differential cross 
sections for the $\gamma p \rightarrow K^+ \Lambda$ reaction. 
The closed circles, open squares and triangles are from the LEPS, 
the SAPHIR~\cite{saphirk+} and the CLAS~\cite{CLAS}, respectively. 
The dot-dashed and dotted curves are the results of the Regge model 
with the $K$- and $K^*$-exchanges and the $K^*$-exchange, respectively, 
obtained by Guidal {\em et al.}~\cite{Guidalnew}. 
The solid curves indicate the result of the mixing models of the Feynman 
diagram and the Regge model~\cite{bennholdcomi}.} 
\end{figure}

The differential cross sections for the $K^+\Lambda$ reaction are shown 
as a function of the total energy comparing with theoretical calculations 
in Fig.~\ref{CrossLambda2}. The dot-dashed and dotted curves are the 
results of the Regge model with the $K$ and $K^*$ exchanges, and only 
$K^*$ exchange, respectively, obtained by Guidal {\em et al.}
~\cite{Guidalnew}. The solid curves indicate the result of the mixing 
models of the Feynman diagram and the Regge model~\cite{bennholdcomi}. 

Mart and Bennhold's model calculation shows a good agreement with 
the LEPS and the CLAS data in all ranges. The resonance-like structure 
at $W$=1.96 GeV is well reproduced by including the missing resonance 
$D_{13}$(1900). The Ghent model calculation also include the 
$D_{13}$(1900) resonance to reproduce the resonance-like structure
~\cite{Janssennew, Janssennew2}. 
The $K$ and $K^*$ exchanges model calculation of the Regge theory 
overestimates the data. The difference between the $K$ and $K^*$
exchanges model, compared with the $K^*$ exchange only model, 
becomes large at 
forward angles because $K$ exchange is dominant at forward 
angles and makes a forward peak for the $K^+\Lambda$ reaction in 
the Regge model. 


\begin{figure}
\includegraphics[height=9.cm]{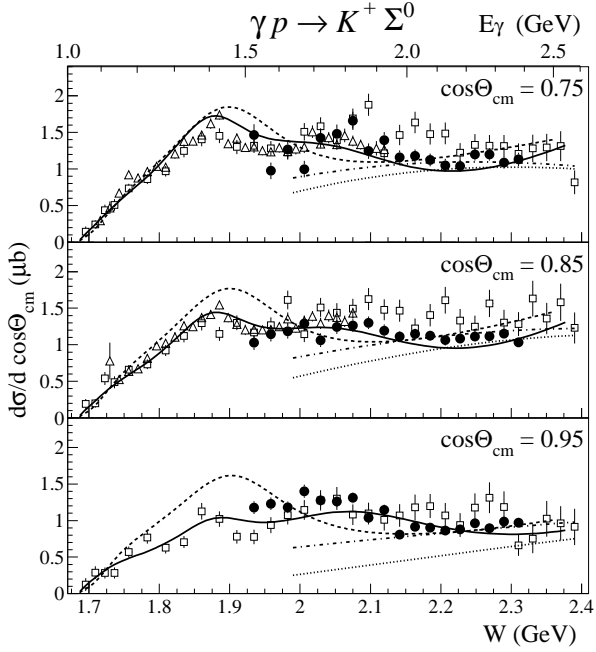}
\caption{\label{CrossSigma2}Energy dependence of differential cross 
sections for the $\gamma p \rightarrow K^+ \Sigma^0$ reaction. 
The closed circles, open squares and triangles are from the LEPS, 
the SAPHIR~\cite{saphirk+} and the CLAS~\cite{CLAS}, respectively. 
Dot-dashed and dotted curves are the result of the Regge model with 
the $K$- and $K^*$-exchanges and the $K^*$ exchange, respectively 
obtained by Guidal {\em et al.}~\cite{Guidalnew}. 
Solid curves indicate the result of the mixing models of Feynman 
diagram and Regge model~\cite{bennholdcomi}. Dashed curves are the 
model calculation by the Ghent group~\cite{corthalscomi}.} 
\end{figure}

The differential cross sections for the $K^+\Sigma^0$ reaction are 
shown as a function of the total energy with the results of the 
theoretical calculations in Fig.~\ref{CrossSigma2}. Dot-dashed and 
dotted curves are the results of the Regge model with the $K$- and 
$K^*$-exchanges, and the $K^*$ exchange, respectively obtained by Guidal 
{\em et al.}~\cite{Guidalnew}. Solid curves indicate the result of 
the mixing models of the Feynman diagram and the Regge model
~\cite{bennholdcomi}. 
Dashed curves are the model calculation by the Ghent group~\cite{
corthalscomi}. 

The $K$ and $K^*$ exchange model underestimates the data at $W<2.15$ GeV  
but shows an agreement at $W>2.15$ GeV where the $t$-channel contribution 
is expected to be dominant. 
In the Regge model, the contribution of the $K$ exchange for the 
$K^+\Sigma^0$ reaction is smaller than that for the $K^+\Lambda$ reaction 
since the coupling constant $|g_{K\Sigma N}|$ is smaller than 
$|g_{K\Lambda N}|$~\cite{Guidal}. 
The difference between the Regge models using $K$ and $K^*$ exchanges,
compared with the model having only $K^*$ exchange, 
is smaller than that for the $K^+\Lambda$ reaction. 

A small enhancement is seen at $W$=2.05 GeV. 
The Ghent isobar-model calculation does not introduce a resonance 
at this region, and underestimates the experimental data. 
Large predicted photon beam asymmetries compared with 
the experimental data may be explained by the absence of 
a resonance in the calculation.
On the other hand, in the Mart-Bennhold's model calculation 
the $P_{31}$(1910) $\Delta$ resonance moves to $W$=2.05 GeV, and 
the measured differential cross sections are reasonably 
reproduced in this energy region. 
The $\Delta^*$ resonance strongly couples to the $K^+\Sigma^0$ channel, 
and a $\Delta^*$ resonance seems to be required to explain the 
enhancement at $W=2.05$ GeV. 

\begin{figure}
\includegraphics[height=13cm]{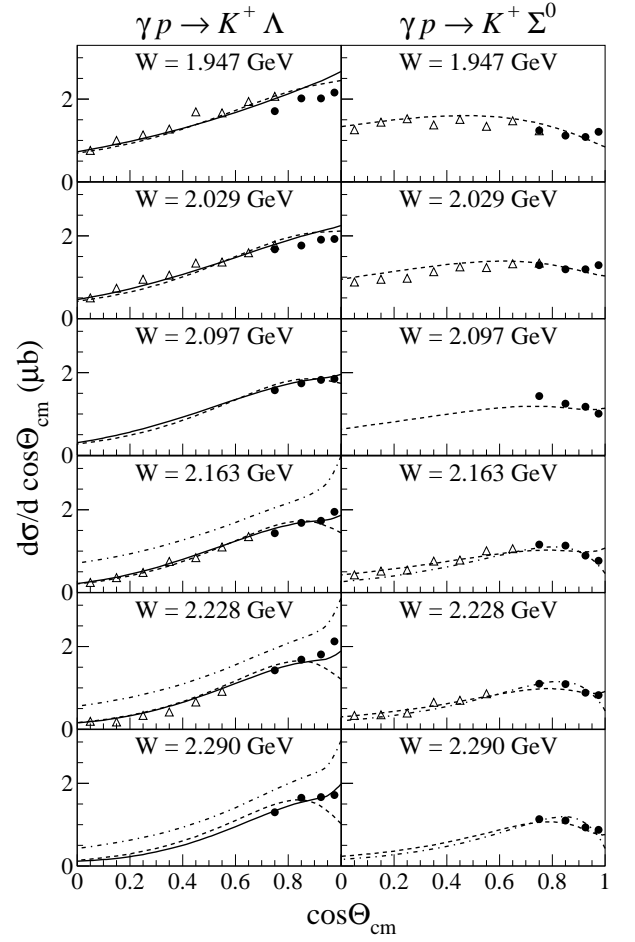}
\caption{\label{CrossAngle2}Angular dependence of differential cross 
sections for the $\gamma p \rightarrow K^+ \Lambda$ (left) and 
$\gamma p \rightarrow K^+ \Sigma^0$ (right) reactions. 
Closed circles are LEPS data and open triangles are CLAS data. 
Dot-dashed, dashed and solid curves are the theoretical calculations  
with the Regge model, the Feynman diagram and the mixing model of 
the Regge model and the Feynman diagram, respectively~\cite{
bennholdcomi}.}
\end{figure}
The differential cross sections are shown as a function of the $K^+$ 
scattering angle in Fig.~\ref{CrossAngle2}. The LEPS data connect 
smoothly to the CLAS data. It is seen that the $K^+\Lambda$ cross 
section increases at forward angles while the $K^+\Sigma^0$ cross 
section decreases except for the low energy regions of $W$=1.947 and 
2.029 GeV. 
The experimental data for the $K^+\Lambda$ and $K^+\Sigma^0$  
reactions are compared with Mart and Bennhold's model calculation 
in Fig.~\ref{CrossAngle2}. The dot-dashed, dashed and solid curves 
are the results of the Regge model, the Feynman diagram and the 
mixing model of the Regge and the Feynman, respectively~\cite{bennholdcomi}. 

The mixing model calculation agrees with the data for the $K^+\Lambda$ 
reaction while the calculation of the Feynman diagram only agrees with 
the data for the $K^+\Sigma^0$ reaction. 
The calculations of the Feynman diagram increase as the scattering angle 
becomes smaller, then they drop at cos$\Theta_{cm}>$0.85 for both reactions.
The model calculation without inclusion of Regge amplitudes cannot explain 
the observed angular distributions for the $K^+\Lambda$ reaction.
The Regge model calculation shows steep increase for the $K^+\Lambda$ 
while it drops for the $K^+\Sigma^0$ at cos$\Theta_{cm}>$0.9.
In the Regge model, the $K$ exchange contribution is large for
the $K^+\Lambda$ but is small for the $K^+\Sigma^0$ at forward angles.
In the high energy data measured at $W>3.2$ GeV at SLAC, the
$K^+\Lambda$ shows a forward peak but the $K^+\Sigma^0$ does not~\cite{SLAC}.
This result was discussed in terms of the dominance of the 
$K$ exchange for the $K^+\Lambda$~\cite{Guidal}.
In our data the same feature is seen at $W$=2.1 to 2.3 GeV.  The
mixing model, which includes the Regge model with a dominant
$K$ exchange contribution, reproduces well the differential
cross sections for the $K^+\Lambda$ reaction.
 
\section{SUMMARY}
The photon beam asymmetries and differential cross sections for the 
$\gamma p \rightarrow K^+ \Lambda$ and $\gamma p \rightarrow K^+ 
\Sigma^0$ reactions have been measured at $E_\gamma$=1.5$-$2.4 GeV 
and at 0.6$<$cos$\Theta_{cm}<$1 by using linearly polarized photons 
at the SPring-8/LEPS facility. 
The photon beam asymmetry data for the $\gamma p \rightarrow K^+ \Lambda$ 
and $\gamma p \rightarrow K^+ \Sigma^0$ reactions have been obtained for  
the first time in this energy range. The sign of the photon beam 
asymmetry has been found to be positive.

We obtained differential cross sections with good statistics at 
forward angles. The present data of the differential cross sections 
are consistent with those obtained by the CLAS Collaboration in the 
overlapping region. The differences of these two data are within the expected 
error. The resonance structure at $W$=1.96 GeV is seen in the $\gamma p 
\rightarrow K^+ \Lambda$ reaction and this is expected to be the same 
structure as one found in the SAPHIR and CLAS data. This bump structure 
can be explained by including a $D_{13}$(1900) resonance. 
A small enhancement has been found at $W$=2.05 GeV in the $\gamma p 
\rightarrow K^+ \Sigma^0$ reaction and the structure is partly reproduced 
by including the $P_{31}$ $\Delta^*$ resonance. The differential 
cross sections for the $\gamma p \rightarrow K^+ \Lambda$ reaction 
rise at forward angles while the cross sections for the $\gamma p 
\rightarrow K^+ \Sigma^0$ reaction drop. This forward peak in the 
$K^+\Lambda$ channel comes from the large contribution of $K$ exchange 
in the $t$-channel. Our data indicate that $K$ exchange is 
dominant in the $\gamma p \rightarrow K^+ \Lambda$ reaction, but not 
dominant in the $\gamma p \rightarrow K^+ \Sigma^0$ reaction. 

\begin{acknowledgments}
We thank the staff at SPring-8 for providing excellent experimental 
conditions during the long experiment. 
This research was supported in part by the Ministry of Education, 
Science, Sports and Culture of Japan, by the National Science Council 
of the Republic of China (Taiwan), and by the National Science Foundation 
(USA).
\end{acknowledgments}


\end{document}